\documentclass[useAMS,usenatbib]{mn2e}

\usepackage{amsmath}
\usepackage{amssymb}
\usepackage[dvipsnames]{xcolor}
\usepackage{graphicx}
\usepackage{graphics}
\usepackage{hyperref}
\usepackage{comment}
\usepackage[autostyle]{csquotes}

\hypersetup{
     colorlinks   = true,
     citecolor    = blue
}

\def\apj{{ ApJ}}
\def\apjl{{ApJL}}

\def\aap{{ A\&A}}
\def\pasp{{ PASP}}

\def\mnras{{ MNRAS}}
\def\araa{{ ARA\&A}}

\def\nat {{ Nature}}

\def\ssr{{ Space Sci. Rev.}}

\def\physrep{{ Physics Reports}}

\def\mr{\mathrm}

\def\log{\mr{log}}

\def\omp{\omega_{\rm p}}

\def\me{m_{\rm e}}

\def\Gams{\Gamma_{\rm s}}
\def\Gamt{\Gamma_{\rm t}}
\def\para{\parallel}

\newcommand{\frb}{{FRB 200428} }
\newcommand{\sgr}{{SGR 1935+2154} }
\newcommand{\tq}{\textquote}

\newcommand{\myemail}{wenbinlu@caltech.edu}

\title[Galactic FRBs]{A unified picture of Galactic and cosmological fast radio bursts}
\author[Lu, Kumar \& Zhang]
  {Wenbin Lu$^1$\thanks{\myemail}, Pawan Kumar$^2$, and Bing Zhang$^3$\\
  $^1$Theoretical Astrophysics, Walter Burke Institute for Theoretical Physics, Mail Code
  350-17, Caltech, Pasadena, CA 91125, USA\\
  $^2$Department of Astronomy, University of Texas at Austin, Austin, TX 78712, USA\\
  $^3$Department of Physics and Astronomy, University of Nevada, Las Vegas, Las Vegas, NV 89154, USA}

\begin{document}
\label{firstpage}
\maketitle

\begin{abstract}
%\frb was spatially associated with the soft gamma-ray repeater \sgr and temporally coincident with one of the X-ray bursts from this magnetar. The volumetric rate of FRB 200428-like events is consistent with the faint end of the cosmological FRB rate function, so this event should provide valuable clues for understanding the entire FRB population. Comparison between the rates of magnetar X-ray bursts and FRBs shows that only a small fraction (roughly $10^{-3}$) of X-ray bursts may be accompanied by FRBs. The coexistence of SGR-related FRBs and cosmological, active FRB repeaters suggests that there might be two distinct formation channels of magnetars to power repeating FRBs.
The discovery of a fast radio burst (FRB) in our galaxy associated with a magnetar (neutron star with strong magnetic field) has provided a critical piece of information to help us finally understand these enigmatic transients. We show that the volumetric rate of Galactic-FRB like events is consistent with the faint end of the cosmological FRB rate, and hence they most likely belong to the same class of transients. The Galactic FRB had an accompanying X-ray burst but many X-ray bursts from the same object had no radio counterpart. Their relative rates suggest that for every FRB there are roughly $10^2$--$10^{3}$ X-ray bursts. The radio lightcurve of the Galactic FRB had two spikes separated by 30 ms in the 400-800 MHz frequency band. This is an important clue and highly constraining of the class of models where the radio emission is produced outside the light-cylinder of the magnetar. We suggest that magnetic disturbances close to the magnetar surface propagate to a distance of a few tens of neutron star radii where they damp and produce radio emission. The coincident hard X-ray spikes associated with the two FRB pulses seen in this burst and the flux ratio between the two frequency bands can be understood in this scenario. This model provides a unified picture for faint bursts like the Galactic FRB as well as the bright events seen at cosmological distances.
\end{abstract}


\section{Introduction}

On April 28, 2020, the Canadian Hydrogen Intensity Mapping Experiment (CHIME, 400-800 MHz) and the Survey for Transient Astronomical Radio Emission 2 (STARE2, 1.3-1.5 GHz) independently detected a fast radio burst (hereafter FRB 200428), which is spatially coincident with the well known Galactic Soft Gamma-ray Repeater (SGR) 1935+2154 \citep{chime20, bochenek20}. The arrival-time difference between these two frequency bands is consistent with dispersive delay due to plasma along the line of sight with dispersion measure $\mr{DM}=332.8\pm0.1\rm \, pc\, cm^{-3}$. The burst had two $\sim 1\rm\, ms$ components separated by about 30 ms as measured by CHIME, the first at lower frequencies (400-550 MHz) and the second at higher frequencies (550-800 MHz). The FRB occurred in a side lobe of CHIME, so its inferred fluence of a few hundred kJy ms may suffer large uncertainty \citep{chime20}. However, STARE2 provided a more accurate fluence measurement of $1.5\rm \, MJy\, ms$ \citep{bochenek20}.

The SGR 1935+2154 was first detected by \textit{Swift} with a burst of $\gamma$-rays \citep{stamatikos14}. Subsequent X-ray follow-up observations identified this source as a magnetar with rotational period $P=3.24\,\rm s$ and characteristic surface dipolar magnetic field $B\simeq 2.2\times10^{14}\rm\, G$ \citep{israel16}. This magnetar has had multiple episodes of outbursts since the initial discovery \citep{lin20a}. \sgr is spatially associated with the supernova remnant G57.2+0.8 \citep{sieber84, gaensler14, surnis16}, which is at a distance between 6.7 and 12.5 kpc from us \citep{kothes18}. We adopt $d\simeq 10\rm\, kpc$ but our results are unaffected by the distance uncertainty.

A hard X-ray burst was detected from \sgr by several instruments including INTEGRAL \citep{mereghetti20}, Insight-HXMT \citep{li20_hxmt}, AGILE \citep{agile_atel}, Konus-Wind \citep{ridnaia20}, and the arrival time is in agreement with that of the FRB after de-dispersion. The X-ray burst had fluence of $7\times10^{-7}\rm\, erg\, cm^{-2}$ in the 1-150 keV range, and the lightcurve in the hardest band (27-250 keV) of HXMT showed two distinct peaks separated by about 30 ms \citep{li20_hxmt}, further confirming the association with the FRB 200428. The isotropic energy ($\nu E_\nu$) ratio between the radio and the hard X-ray bands is $\sim 3\times10^{-5}$.

Understanding the origin of FRBs --- mysterious bright millisecond-duration radio flashes first discovered about a decade ago \citep{lorimer07} --- has been a major scientific goal of many current or future telescopes, such as Parkes \citep{thornton13, bhandari18}, Arecibo \citep{spitler16}, UTMOST \citep{caleb17}, ASKAP \citep{shannon18}, CHIME \citep{chime_repeaters}, FAST \citep{li13}, and DSA \citep{ravi19}. Before the discovery of FRB 200428, 
%it has been believed that all FRBs are of extragalactic origin, as indicated by the few sources with identified host galaxies
all localized FRBs were from cosmological distances \citep{chatterjee17, bannister19, prochaska19, ravi19, marcote20}. Even with precise localizations of these events in their host galaxies,
%However, at cosmological distances, even with the precise localization down to a small region in the host galaxy, 
it is so far inconclusive what the progenitors of FRBs are and by what process the powerful radio emission is generated. Many ideas have been proposed \citep[see][for recent reviews]{katz18_review, petroff19, cordes19}. They fall into two general categories: (1) emission within the magnetosphere of a neutron star (NS), and (2) emission from a relativistic outflow which interacts with the surrounding medium at large distances from the NS or black hole.

The isotropic specific energy of FRB 200428, $E_\nu \sim 2\times 10^{26}\rm\, erg\, Hz^{-1}$ (for a distance of 10 kpc), is about a factor of $\sim$30 lower than the faintest burst detected from FRB 180916 at cosmological distances \citep{marcote20} but exceeds that of the brightest known giant radio pulses from NSs by four orders of magnitude. Apart from this energetic argument, we provide further evidence based on volumetric rate (in \S\ref{sec:rate}) that \frb belongs to the faint end of the cosmological FRB population. Therefore, the detection of FRB 200428 in the Milky Way provides an extraordinary opportunity to understand the FRB phenomenon, in the following three major aspects: (1) strongly magnetized NSs or magnetars can make FRBs \citep[as advocated by many authors, e.g.,][]{popov10, kulkarni14, katz16b, kumar17, lyubarsky14, beloborodov17, metzger19, wadiasingh19}, (2) the associated X-ray emission (and future identifications of other counterparts) provides valuable clue for the emission mechanism, (3) the close proximity may allow us to disentangle many of the propagation effects from the intrinsic emission properties.

% The isotropic specific energy of FRB 200428, $E_\nu \gtrsim 3\times 10^{26}\rm\, erg\, Hz^{-1}$ (for a distance of 12.5 kpc), is about a factor of 20 lower than the faintest burst detected from FRB 180916 at cosmological distances \citep{marcote20} but exceeds that of the brightest giant pulses by 6 orders of magnitude. 
% , which corresponds to specific energy $E_\nu \gtrsim 3\times 10^{26}\rm\, erg\, Hz^{-1}$ and $\nu E_\nu \gtrsim 4\times10^{35}\rm\, erg$ (and luminosity $\nu L_\nu \simeq 10^{38}\rm\, erg\, s^{-1}$) for a fiducial distance of 12.5 kpc.

This paper aims to explore the implications of FRB 200428. In \S\ref{sec:rate}, we compare the rate of FRB 200428-like events with that of the cosmological FRB population. In \S\ref{sec:progenitor}, we compare \sgr with the sources of other actively repeating FRBs and discuss how they may be understood in a general framework of the magnetar progenitors from different formation channels. In \S\ref{sec:link}, we compare the rates of magnetar X-ray bursts and FRBs, and discuss the physical link between them. Finally, we closely examine the possible emission mechanisms for FRB 200428 and for other FRBs in \S\ref{sec:emission_mechanism}. A brief summary is provided in \S\ref{sec:summary}. We use the widely adopted, convenient, subscript notation of $X_n \equiv X/10^n$ in the CGS units.

% the first known multi-wavelength counterpart of an FRB. The isotropic energy $\nu E_\nu$ ratio between the radio and the hard X-ray bands is $\gtrsim 3\times10^{-5}$. 

% At UTC 14:34:24.5, Insight-HXMT detected a hard X-ray burst with fluence $6.8\times10^{-7}\rm\, erg\, cm^{-2}$ in the range 1--150 keV and duration $\sim0.5\,$s. INTEGRAL also detected the same X-ray burst of fluence $3.9\times 10^{-7}\rm \, erg\, cm^{-2}$ in the range 25--80 keV. This burst was also detected by Konus-Wind which gives a fluence of $7.6\times10^{-7}\rm\, erg\, cm^{-2}$ in the 20--250 keV range. Fits to the Insight-HXMT and Konus-Wind data sets both yield an exponential spectral cutoff beyond $E_{\rm p}\sim 70\,$keV. The Konus-Wind fluence corresponds to energy $E_{\rm X}\simeq 1.4\times10^{40}\rm\, erg$. The arrival time of the X-ray burst precisely agrees with the de-dispersed FRB arrival time at infinite frequency, which confirms the association. The energy ratio between the radio and hard X-ray bands is $\gtrsim 3\times 10^{-5}$.

% , corresponding to energy $\sim 7\times10^{39}\rm \, erg$. The time UTC 14:34:24 for the X-ray burst precisely agrees with the  Then the energy ratio between radio and hard X-ray bands is 

\section{FRB volumetric rate density}\label{sec:rate}
\begin{figure}
\centering
\includegraphics[width=0.47\textwidth]{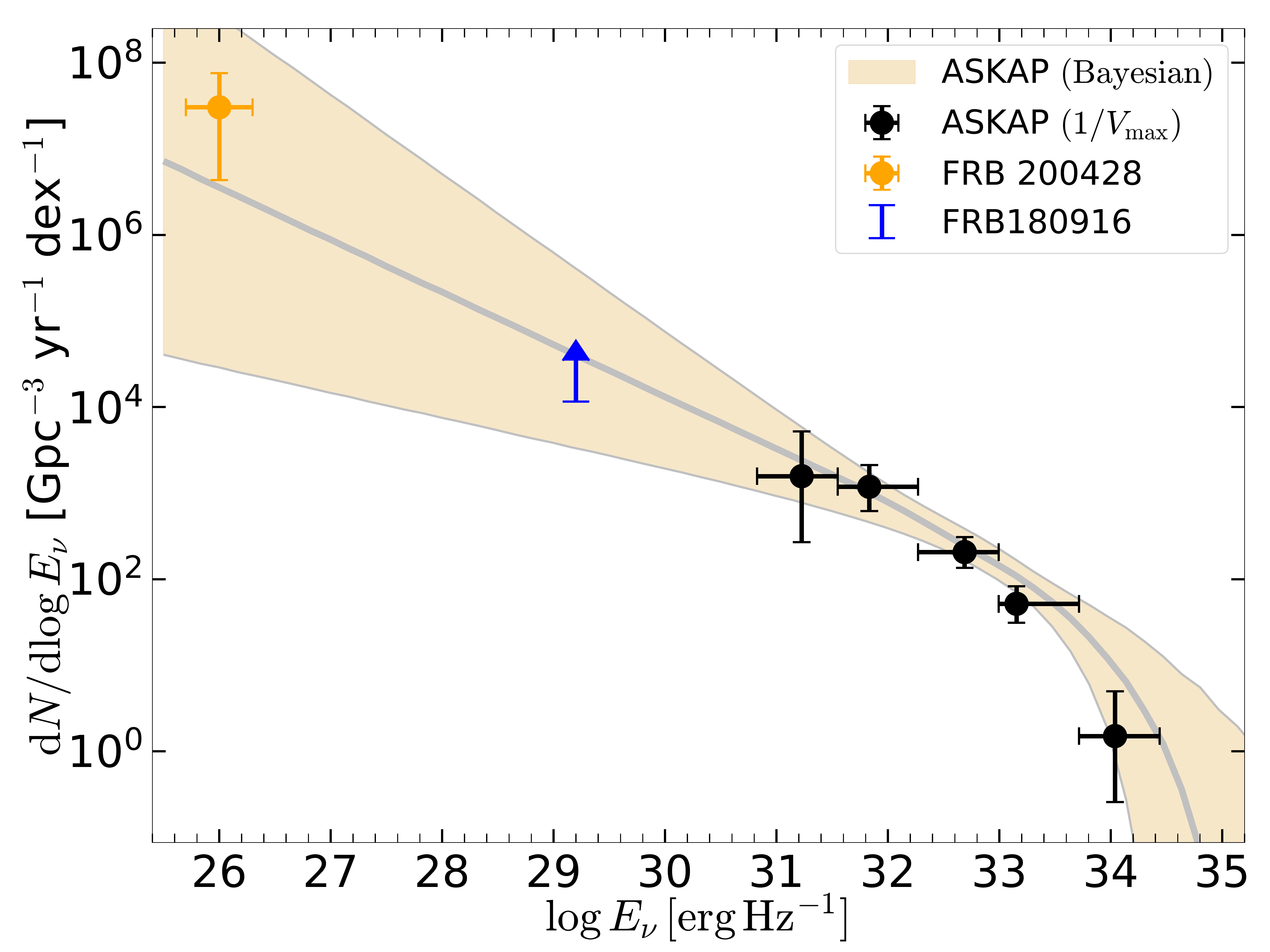}
\caption{The volumetric rate at the faint end as inferred from \frb \citep[orange point with 68\% C.L. Poisson errors,][]{bochenek20}, as compared to the rate at the bright end inferred from the ASKAP sample (the shaded region). The silver lines mark the 16\% ($-1\sigma$), 50\% (median), 84\% ($+1\sigma$) percentiles based on the Bayesian posterior shown in Fig. 4 of \citet{lu19a} and evaluated at redshift $z=0.3$ (where the ASKAP constraints are the strongest). We do not expect significant rate evolution between $z=0.3$ and the local Universe at $z=0$. The black points (with 68\% C.L. Poisson errors) are from an independent analysis of the ASKAP sample based on the classical $1/V_{\rm max}$ estimator \citep{schmidt68}. The blue arrow shows the 90\% C.L. lower limit for the contribution to the total volumetric rate density by FRB 180916 \citep{chime20-period}, although this is measured at $\sim0.6\,$GHz rather than 1.4 GHz.
}
\label{fig:rate_density}
\end{figure}

Based on a single detection in about one year of STARE2 operation \citep{bochenek20_instrument}, we roughly estimate the Galactic FRB rate to be $\sim 10\rm\, yr^{-1}$ above specific energy of $E_\nu\sim 5\times 10^{25}\rm\, erg\,Hz^{-1}$ (the detection threshold energy for a distance of 10 kpc). This leads to a volumetric rate of $\sim 10^8\rm\, Gpc^{-3}\, yr^{-1}$ \citep[see][for a more detailed calculation]{bochenek20}. This should be compared with the bright end of rate density distribution $R\sim 10^{2.6}\rm\, Gpc^{-3}\, yr^{-1}$ above $E_\nu = 10^{32}\rm\, erg\, Hz^{-1}$ as measured by ASKAP also at 1.4 GHz \citep{shannon18, lu19a}. We find the slope for the cumulative rate distribution to be $\beta \simeq \Delta \log R/\Delta \log E_\nu \simeq 0.8$, which is insensitive to Poisson error of a factor of a few. This agrees with the slope of the rate distribution found within the (small) ASKAP sample $0.3\lesssim \beta\lesssim 0.9$ \citep{lu19a} as well as the joint analysis of the Parkes and ASKAP samples $0.5\lesssim \beta \lesssim 1.1$ \citep{luo20}. This agreement suggests that \frb contributes a significant fraction of the FRB rate density at the faint end near $E_\nu\sim 10^{26}\rm\, erg\, Hz^{-1}$, as illustrated by Fig. \ref{fig:rate_density}. Combining this with the fact that the specific energy of \frb is only a factor of $\sim$30 below the faintest known cosmological FRB \citep{marcote20}, we conclude that the magnetar nature of the progenitor and emission mechanism of \frb is likely representative of the whole FRB population.

\section{Nature of FRB progenitors}\label{sec:progenitor}

The number density of galactic magnetars, \sgr being one of them, is of the order $3\times10^8\rm\, Gpc^{-3}$. The progenitors of highly active repeaters like FRB 180916 \citep{chime_repeaters, marcote20} are much rarer in the Universe with a number density of $\sim7$--$700\rm\, Gpc^{-3}$, which is estimated by the expectation number between 0.05 and 4.7 \citep[90\% C.L.,][]{gehrels86} of such repeaters in about half of the sphere (visible by CHIME) within $150\rm\, Mpc$. If these active repeaters are also powered by magnetars, they must belong to a type of ``active magnetars'' not seen in the Milky Way. If one assumes that all active magnetars will evolve to normal magnetars over time by reducing the bursting rate, the volume density of these active magnetars' ``descendants'' would be at most a factor of $\sim 10^4/30 \sim 300$ times greater than the volume density of active magnetars, where $10^4$ yr is the typical age of SGR 1935+2154-like galactic magnetars and $\sim$$30\,\rm yr$ is a conservative estimate of the characteristic age of active magnetars. This gives a volume density of $\sim 2\times 10^3$--$2\times 10^5 \rm\, Gpc^{-3}$ of these descendants, still 3--4 orders of magnitude smaller than the galactic magnetar volume density. The discrepancy is even larger if the characteristic age of active magnetars is longer than 30 yr. This deficit cannot be fully reconciled by reasonable beaming correction\footnote{Here, beaming correction is given by the \textit{total} beaming factor, which is defined as the fraction of the whole $4\pi$ sky occupied by the union of the solid angles spanned by all bursts from a given source. Due to the star's rotation, the \textit{total} beaming factor is typically substantially larger than that for individual bursts.}, because we would not have seen FRB 200428 from one of the $\sim$30 magnetars in our galaxy if the average beaming fraction is $\ll 1/30$. We can then draw the conclusion that ``active magnetars'' and SGR 1935+2154-like galactic magnetars must be two distinct populations \citep[as also suggested by][]{margalit20_galacticFRB}.

One possibility is that the progenitors of FRB 180916 \citep[or FRB 121102,][]{spitler16} may be produced from rare, extreme explosions such as long gamma-ray bursts (LGRBs), superluminous supernovae \citep[SLSNe, e.g.,][]{metzger17}, or NS mergers \citep[e.g.,][]{margalit19,wang20}, so that they have relatively short (e.g. millisecond) periods at births \citep{usov92,zhang01, metzger11}. These magnetars likely stored more toroidal magnetic energy inside the star which provides a larger energy reservoir to power bursting activities \citep[e.g.,][]{thompson93}. In contrast, Galactic magnetars were likely born with a more moderate initial spin, as evidenced by the limited energy in their surrounding supernova remnants \citep[e.g.,][]{vink06}. These magnetars may store less toroidal magnetic energy inside the star and are relatively less active compared with their active cousins. The possible dichotomy of FRB magnetar progenitor is consistent with the host galaxy data of the localized FRBs \citep{li20}: whereas FRB 121102 has a host galaxy similar to that of LGRBs or SLSNe \citep{tendulkar17,metzger17,nicholl17}, other four hosts resemble the Milky Way galaxy that hosts regular magnetars \citep{bannister19,ravi19,prochaska19,marcote20}.

%\note{[--- discuss the number density of normal magnetars $3\times 10^{8}\rm\, Gpc^{-3}$ is much higher than that of active repeaters like FRB 190816 or FRB 121102 ($\lesssim 500\rm\, Gpc^{-3}$, there is no such objects in the Milky Way). This cannot be due to evolution only, which at most gives a factor of 300 as given by the ratio of the ages of the two. A small fraction of magnetars must be born with extreme properties to allow very active repetition. ]}

% \subsection{Beaming}
% Take the energy dependent rate of SGR bursts and scale by $10^{-4}$ to obtain the volumetric beaming-corrected FRB rate, if all SGR bursts are associated to FRBs. This gives a beaming factor of $f_{\rm b}\sim 10^{-2}$--$10^{-3}$.

% Alternatively, it is possible that only a small fraction of SGR bursts are accompanied by FRB emission.

% \subsection{Repeating rates}

\section{Link between X-ray and radio emission}\label{sec:link}
The hard X-ray burst associated with \frb was one of the numerous X-ray bursts that SGRs generate during their active periods. The ratio of the energy release in the radio and X-ray bands is $f_{\rm r}\sim 3\times10^{-5}$. In the following, we discuss the implications on the physical link between emission in these two bands and possible beaming of the radio emission.

The Galactic SGR X-ray burst rate is of order $0.1\rm\, yr^{-1}$ (volumetric rate $\sim 2\times 10^{6}\rm\, Gpc^{-3}\rm\, yr^{-1}$) above $E_{\rm x}=10^{44}\rm\, erg$, and the energy dependence has a similar power-law form as that for FRBs \citep[e.g.,][]{ofek07, kulkarni14, beniamini19}. For a fiducial value of the radio-to-X-ray flux ratio $f_{\rm r}=10^{-4}f_{\rm r,-4}$ to connect the rates of X-ray bursts to FRBs \citep{chen20}, $E_{\rm x}=10^{44}\rm\, erg$ corresponds to FRB specific energy of $E_\nu\simeq 10^{31}f_{\rm r,-4}\rm\,erg\, Hz^{-1}$ (for $1\rm\,GHz$ bandwidth), above which the volumetric rate is $\sim 3\times 10^{3} f_{\rm r,-4}^{-0.8} \rm\, Gpc^{-3}\, yr^{-1}$ \citep{lu19a, luo20}. We see that only a small fraction ($10^{-3}$ to $10^{-2}$) X-ray bursts may be associated with FRBs. This also agrees with the fact that 29 of the X-ray bursts from \sgr had concurrent observations by FAST but no radio signal was detected down to fluence limit of $\sim$10~mJy~ms \citep{lin20b}. 

This small fraction of association may be explained by (a combination of) the following two possible reasons. The first is that most X-ray bursts are accompanied by an FRB but the radio emission is highly beamed, with a beaming fraction of $\Omega_{\rm frb}/4\pi \sim 10^{-3}$--$10^{-2}$. This may be realized if FRBs are only generated along magnetic field lines near the poles. The second explanation is that only a small fraction of X-ray bursts may be physically associated with FRB but in each association the FRB beaming fraction is order unity.
% This may be realized if only special regions (e.g. polar caps) near the magnetar surface may allow particles to flow to large distances and generate FRBs \citep[such as in the models of][]{lyubarsky14, beloborodov17, metzger19}. A combination of these two explanations is also possible. 

% Most SGR X-ray bursts are not associated with FRBs. This can be seen from two different arguments. First, if every X-ray burst is associated with an FRB with fixed energy ratio $f_{\rm r}\sim 10^{-4}$, then the SGR X-ray burst rate ($0.2\rm\, galaxy^{-1}\, yr^{-1}$ above $10^{44}\rm\,erg$) gives FRB volumetric rate of $\sim 2\times10^{7}\rm\, Gpc^{-3}\, yr^{-1}$ above $10^{31}\rm\, erg\, Hz^{-1}$, which is a factor of $10^3$ higher than the actual FRB rate. Second, for \sgr, many (a few tens) X-ray bursts are detected with simultaneous radio coverage but non-detection.

% More than a few tens of $\gamma$/X-ray bursts at the level of the hard X-ray burst associated with \frb have been detected by various instruments, e.g. \textit{Swift} BAT, \textit{Fermi} GBM, AGILE, and NICER. No FRBs were found coincident with the other bursts, even though STARE2 is sensitive to fainter FRBs down to $\sim 300\rm\, kJy\, ms$. This suggests that either most $\gamma$/X-ray bursts are not intrinsically related to FRB emission or the latter is strongly beamed in random directions such that the observer only sees a small fraction of all FRBs. This further disfavors the synchrotron maser model in which the relativistic ejecta in each flare is quasi-isotropic and the overall beaming factor should be order unity.

In the next section, we discuss the implications of the association between FRBs and magnetar X-ray bursts on the coherent emission mechanism.
\begin{figure*}
\centering
\includegraphics[width=0.5\textwidth]{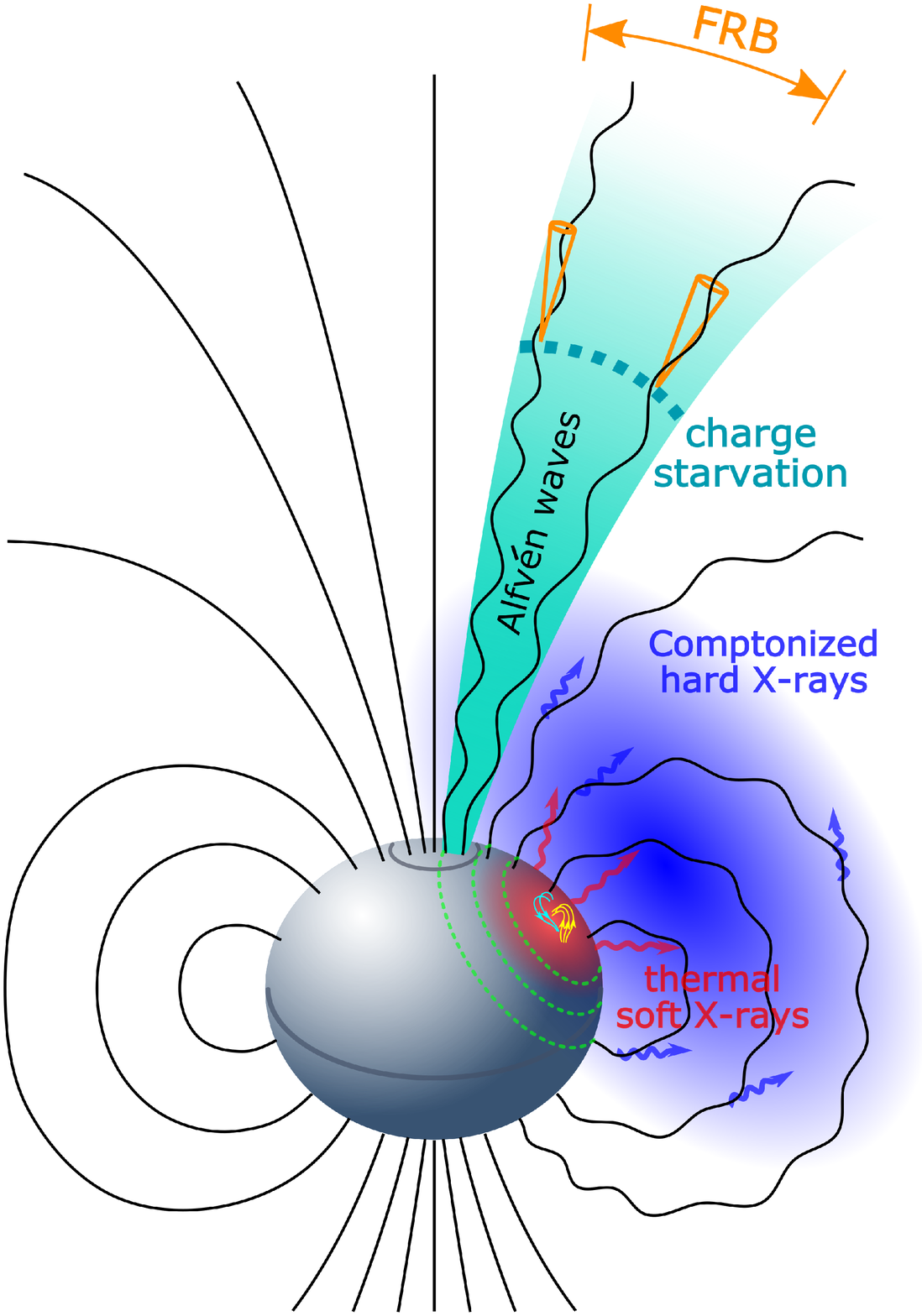}
\hspace{1.5cm}
\includegraphics[width=0.25\textwidth]{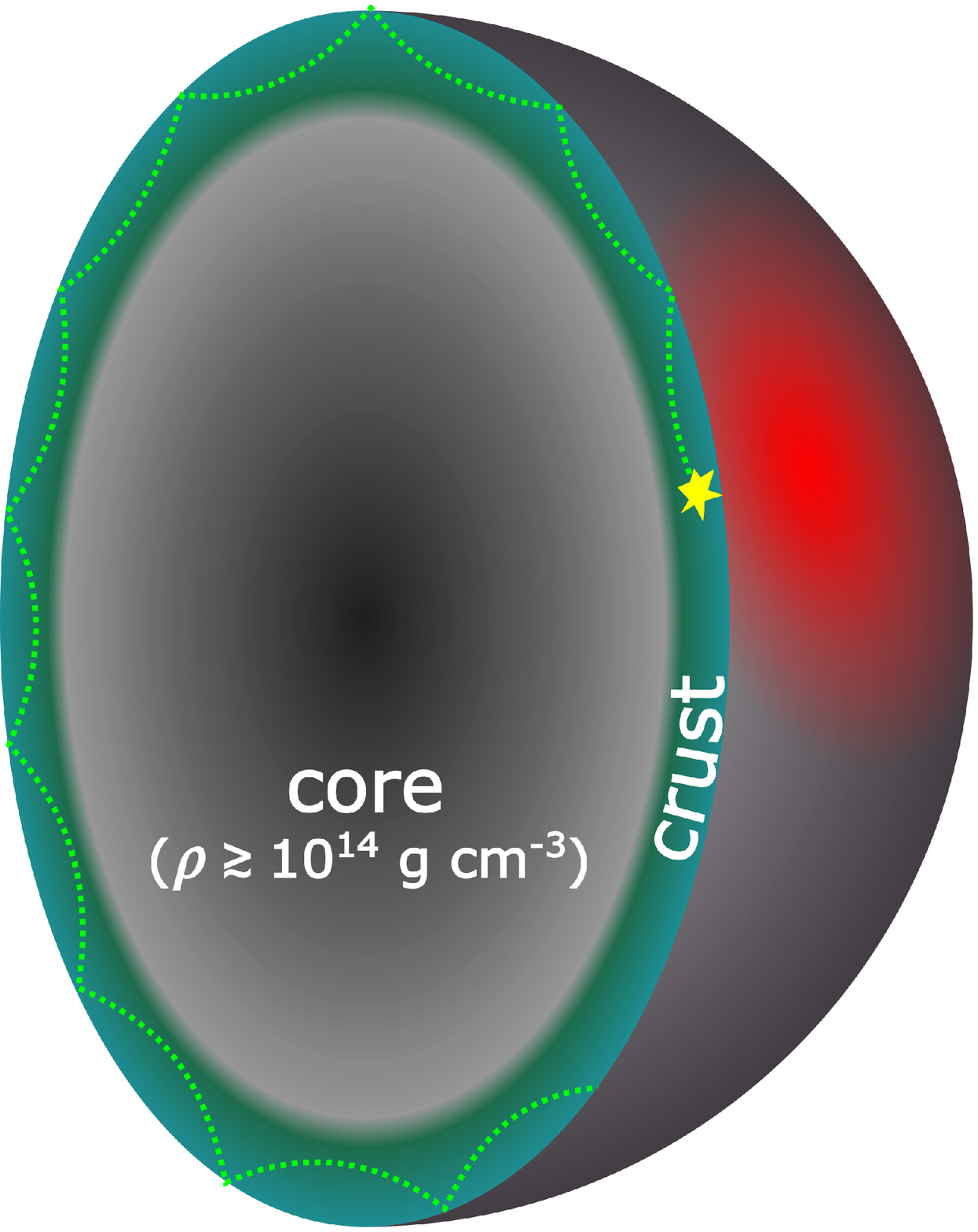}
\caption{Sketch of the model described in this paper. \textit{Left panel}: Sudden magnetic energy dissipation heats up the NS surface and generate $\mr{e}^{\pm}$ pair fireball which is trapped by the closed field lines. X-rays are produced by the heated surface (red shaded region) and then inverse Compton scattered by $\mr{e}^{\pm}$ pairs (blue shaded region) in the magnetosphere to higher energies. The disturbance spreads across the NS surface (green dashed circles) and launches Alfv{\'e}n waves (shown as wiggles with exaggerated amplitude) which propagate along magnetic field lines. Near the magnetic poles, Alfv{\'e}n waves can reach distances much larger than the NS radius where the charge density is too low to sustain the plasma current associated with the wave (marked by a teal dashed curve). This is because the plasma density in the magnetosphere drops rapidly with the distance to the NS. As a result of charge starvation, a strong electric field parallel to the background magnetic field develops, and charge clumps are accelerated to high Lorentz factors and coherently produce curvature emission in the radio band (marked as orange cones). In this picture, the FRB emission is narrowly beamed into the region spanned by the orange arrows, whereas the X-rays are visible from a large fraction of the sky. The double radio pulses seen in \frb are produced by two separate eruptions, which also enhances Comptonization and gives rise to the two hard X-ray peaks. \textit{Right panel}: Crustal deformations due to sudden magnetic energy release excite shear mode oscillations. The shear wave propagates along the crust, and when it reaches the NS surface, a fraction of energy is transmitted into the magnetosphere as Alfv{\'e}n waves and the rest is reflected back into the crust. The FRB duration is given by shear wave propagation delay between different paths, $t_{\rm frb}\sim 1\rm\, ms$ for typical wave speed $v_{\rm s}\sim 0.01c$.
% In this scenario, the FRB duration is determined by wave propagation inside the crust, $t_{\rm frb}\sim R_{\rm ns}/v_{\rm s}\sim 3\rm\, ms$ for typical wave speed $v_{\rm s}\sim 0.01c$.
% High-frequency shear waves are quickly transmitted to the magnetosphere in the closed field line region, whereas low-frequency waves reflect off the surface many times and reach the magnetic poles where they can efficiently launch Alfv{\'e}n waves.
}
\label{fig:Alfven_waves}
\end{figure*}

\section{Emission mechanism}\label{sec:emission_mechanism}

\begin{figure*}
\centering
\includegraphics[width=0.8\textwidth]{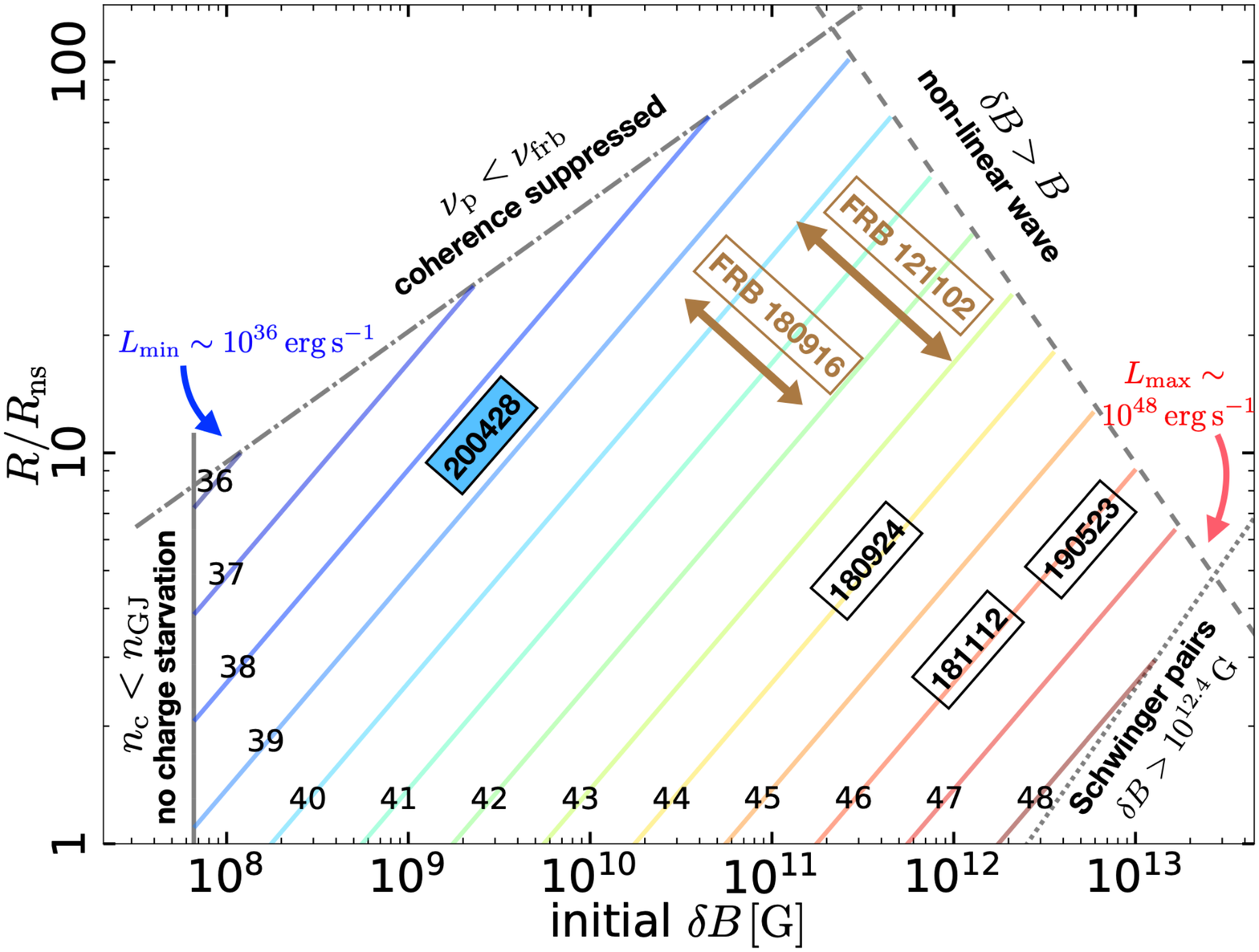}
\caption{Physically allowed initial Alfv{\'e}n wave amplitude $\delta B$, charge starvation radius $R$ (in units of NS radius $R_{\rm ns}$), and FRB luminosity for the model described in this paper. The boundaries of the parameter space are given by the following constraints (shown by the grey lines): (1) the critical density $n_{\rm c}(R)$ must be greater than the Goldreich-Julian density $n_{\rm GJ}$ for charge starvation to be possible (solid), (2) the plasma frequency $\nu_{\rm p}$ must exceed the FRB frequency $\nu_{\rm frb}$ so as to allow charge clumps of size $\ell_\para\sim \lambda_{\rm frb}$ (dash-dotted), (3) the wave amplitude at the critical radius $\delta B(R)$ is less than the background magnetic field $B$ so the wave remains linear (dashed), (4) the wave amplitude at the critical radius must not exceed $\sim 5\%$ of the quantum critical field strength to avoid rapid production of Schwinger pairs (dotted). The solutions for different FRB luminosities from $10^{36}$ to $10^{48}\rm\, erg\, s^{-1}$ lie on the colored lines (with $\mr{log}\,L_{\rm frb}\rm \,[erg\,s^{-1}]$ marked on each line). Since the charge density profile in the NS magnetosphere is poorly understood, our current model cannot provide a unique solution. We mark the localized sources with known (ranges of) luminosities in boxes, with the repeaters (FRB 121102 and 180916) in brown. The parameters used for this plot are: transverse wavelength $\lambda_{\perp}=10^{4}\rm\, cm$, surface magnetic field $B_{\rm ns}=3\times 10^{14}\rm\,G$, spin period $P=3\rm\,s$, magnetic colatitude of the field line $\theta=10^{-1.5}\rm\,rad$, and observing frequency $\nu_{\rm frb}=1\rm\,GHz$.
}
\label{fig:constraints}
\end{figure*}

Models for the generation of FRB coherent radio emission can be divided into two broad classes based on the distance from the NS where they operate. The first class consists of the \tq{far-away} models where relativistic ejecta from a neutron star (or black hole) dissipates its energy at large distances by interacting with the circum-stellar medium (CSM) and the radio emission is generated by a maser process \citep{lyubarsky14, waxman17, beloborodov17, beloborodov19, metzger19, margalit20}. The second class are the \tq{close-in} models which describe that the coherent processes occur within the magnetosphere of a neutron star \citep{pen15, cordes16, lyutikov16, kumar17, zhang17, lu18, yang18, kumar20, lyubarsky20, wang20_magnetosphere}. These two general classes of models have very different predictions regarding the FRB temporal and spectral properties, and multiwavelength counterparts. In the Appendix, we present a detailed analysis of the \tq{far-away} models and show that they face a number of difficulties explaining the available radio and X-ray data for FRB 200428.

In this section, we focus on the generation of FRB radiation in the magnetosphere of a magnetar. An additional motivation for our consideration of this model is that at least some FRBs show very rapid variability time as short as tens of micro-seconds \citep{farah18, prochaska19, cho20}, which corresponds to the light-crossing time of a few km and suggests that the radiation might be produced close to a neutron star\footnote{The transverse size of the source and the distance from the compact object is larger when the radiation is produced in a relativistic outflow moving toward the observer with a Lorentz factor $\gamma$ by a factor $\gamma$ and $\gamma^2$ respectively \citep[e.g.,][]{katz19_energetics}.}.

% \subsection{Emission from within the magnetosphere --- coherent curvature model}\label{sec:magnetosphere}
% one of the motivations: rapid variability time (down to $\rm \mu s$).

% In the previous subsection we presented an analysis of the possibility that the FRB emission is produced at distances beyond the light cylinder, and found a number of difficulties with this scenario. This motivates us to investigate the generation of FRB radiation in the magnetosphere of the magnetar \citep{kumar17, lu18, yang18, kumar20}. An additional motivation is that at least some FRBs show very rapid variability time as short as tens of micro-seconds \citep[down to what our instruments allow,][]{farah18, prochaska19}, which corresponds to a light-crossing time for 10 km.

The basic scenario we suggest is that a disturbance emanating from the NS surface spreads through the magnetosphere. The dissipation of the energy near the surface in the closed field line region produces X-ray emission. The disturbance propagating to distances much larger than the NS radius, above the magnetic poles, is converted into coherent radio waves (Fig. \ref{fig:Alfven_waves}). 

Let us first consider that the energy in the outburst near the surface of the NS is carried by a beam of $e^{\pm}$ pairs of isotropic equivalent luminosity $L_{\rm b}$ and Lorentz factor $\gamma_{\rm b}$. The $e^\pm$ number density at distance $R=10^8R_8\rm\, cm$ from the NS in the beam comoving frame is given by $n_{\rm b} \sim 3\times10^{16}\mr{\, cm^{-3}}\, R_8^{-2}\gamma_{\rm b}^{-2}$ for $L_{\rm b} = 4\pi R^2 \gamma_{\rm b}^2 n_{\rm b} \me c^3 \sim 10^{38}\rm\, erg\, s^{-1}$, which is the minimum particle beam luminosity so as to generate the observed radio flux of FRB 200428. This corresponds to a plasma frequency of $\nu_{\rm p} \sim 10^{13} R_{8}^{-1}\mr{\, Hz}$ in the observer frame. Moreover, the cyclotron frequency is $\nu_{\rm B}\sim 3\times10^{15} R_8^{-3}\mr{\, Hz}$ for surface dipolar magnetic field strength of $B_{\rm ns}=10^{15}\,\rm G$. Most maser processes resulting from an interaction between highly relativistic beam of particles and mildly or sub-relativistic plasma produce radiation near the plasma frequency or the appropriately Doppler shifted cyclotron frequency. The estimates for these frequencies show the difficulty for the particle beam based class of maser models to produce GHz radiation with the observed luminosity of FRB 200428.

We consider another possibility, that the energy released near the polar region of the NS is carried by magnetic disturbances -- Alfv{\'e}n waves -- which damp far away from the surface, but well inside the light cylinder, and produce radio waves \citep{kumar20}. Let us consider that the amplitude of the Alfv{\'e}n wave at the NS surface is $\delta B$ and its transverse wavenumber is $k_\perp = 2\pi/\lambda_\perp$, where $\lambda_\perp$ is the wavelength perpendicular to the NS magnetic field. Both $\delta B$ and $k_\perp$ decrease with radius as $R^{-3/2}$ as the wave packet follows the curved magnetic field lines and fans out such that its transverse size increases as $R^{3/2}$. The wave becomes charge starved at a radius $R$ where the plasma density is below the critical density
\begin{equation}
  n_c = {|\nabla\times {\bf \delta B}|\over 4\pi q} \approx {k_\perp \delta B\over 4\pi q} \simeq (1\times10^{12}\mr{\,cm^{-3}})\, R_7^{-3} {\delta B_{10}\over \lambda_{\perp4}},
\end{equation}
% The wave becomes charged starved when
% \begin{equation}
%     |\nabla\times {\bf \delta B}| \approx k_\perp \delta B > 4\pi q n,
% \end{equation}
% The charge starvation radius $R$ can be obtained by comparing the plasma density $n$ to the critical density
% \begin{equation}
%   n_c = {k_\perp \delta B\over 4\pi q} \simeq (1\times10^{15}\mr{\,cm^{-3}})\, R_6^{-3} {\delta B_{10}\over \lambda_{\perp4}},
% \end{equation}
where $q$ is electron charge, $\delta B_{10} = \delta B/10^{10}\rm\, G$ and $\lambda_{\perp 4} = \lambda_\perp/10^4{\rm\, cm}$ are measured at the NS surface. 
% Our fiducial value of the transverse wavelength is motivated by the magnetic energy consideration $\lambda_{\perp}\sim (8\pi E/B_{\rm ns}^2)^{1/3} \sim 10^4\mr{\, cm}\, E_{40}^{1/3} B_{\rm ns,15}^{-2/3}$, where $E = 10^{40}E_{40}\rm\, erg$ is the total energy released in each event.

When the wave arrives at the charge starvation radius $R$, a strong electric field develops along the background magnetic field and accelerates clumps of particles that were formed due to two-stream instability associated with the Alfv{\'e}n wave current density. These particle clumps move along curved field lines and produce coherent curvature radiation. The clumps that form due to two-stream instability have a broad spectrum of longitudinal sizes $\ell_\parallel\lesssim c/\nu_{\rm p}$ ($c$ being the speed of light), and radio emission is generated by those ones with $\ell_\para\simeq \lambda_{\rm frb}/2 =15\nu_9^{-1}\rm\, cm$. The number of particles that can radiate coherently is $N_{\rm coh}\simeq \pi n_{\rm c} \ell_\para \ell_\perp^2$, where the transverse size is given by $\ell_\perp \simeq \sqrt{R\lambda_{\rm frb}}$ such that the photon arrival time does not differ by more than half an FRB wave period. This choice of $\ell_\perp$ is because the other two relevant length scales --- the Alfv{\'e}n transverse wavelength $\lambda_\perp$ and the causal length $R/\gamma$ --- are typically much longer than $\sqrt{R\lambda_{\rm frb}}$.
% $\mr{min}[\sqrt{R\lambda_{\rm frb}/2}, \lambda_\perp/2, R/\gamma]$. Here, $\sqrt{R\lambda_{\rm frb}/2}$ guarantees that the photon arrival time does not differ by more than half a (FRB) wave period, $\lambda_\perp$ is the Alf{\'e}n transverse wavelength, and $R/\gamma$ is the size of the causally connected region.
The clump Lorentz factor $\gamma$ is related to the characteristic frequency of curvature emission $\nu = 3\gamma^3 c/(4\pi R_{\rm B})$ and the curvature radius of magnetic field lines $R_{\rm B}$,
\begin{equation}
    \gamma \simeq 240\, (\nu_9 R_{\rm B,8})^{1/3}.
\end{equation}
The total luminosity is $N_{\rm coh}^2$ times the curvature luminosity $L_{\rm curv}\simeq 16\gamma^8 q^2c/3R_{\rm B}^2$ from an individual particle, provided that the observer is located within the relativistic beaming cone (of angular size $\sim \gamma^{-1}$), so we obtain
\begin{equation}\label{eq:Lfrb}
    L_{\rm frb} \simeq 7\times10^{39}\mr{\, erg\, s^{-1}}\, {(\delta B_{10}/\lambda_{\perp4})^2 \over R_7^{11/3} \theta_{-1.5}^{2/3} \nu_9^{4/3}},
\end{equation}
where we have denoted the magnetic colatitude of the field line on the NS surface as $\theta=10^{-1.5}\theta_{-1.5}\rm\, rad$ and the corresponding curvature radius for a dipolar geometry is $R_{\rm B}\simeq 0.8(R_{\rm ns}/\theta) (R/R_{\rm ns})^{1/2}$.

% The number of particles that can radiate coherently is $N_{\rm coh} \simeq \pi n_c \gamma^2(\lambda_{\rm frb}/2)^3$, where $\gamma$ is the clump Lorentz factor and is related to the characteristic frequency $\nu = 3\gamma^3 c/(4\pi R_{\rm B})$ and curvature radius of magnetic field lines $R_{\rm B}$. The transverse size of the emitting region visible to the observer is $R/\gamma$, which contains $(R/\gamma)^2/(\gamma \lambda_{\rm frb}/2)^2$ mutually uncorrelated coherent patches. \note{[there's a problem with $R/\gamma$, because that will give very smooth FRB frequency spectrum $\delta \nu/\nu \sim 1$. ]} Collecting all these pieces together, we find that the total curvature luminosity is
% \begin{equation}
%     L_{\rm frb} \simeq {\pi^2 n_{\rm c}^2\gamma^8 \lambda_{\rm frb}^4 q^2 c R^2 \over 3R_{\rm B}^2}\simeq 2\times10^{43} \mr{\,erg\, s^{-1}} {(\delta B_{10}/\lambda_{\perp4})^2 \over R_6^{11/3} \theta_{-2}^{2/3} \nu_9^{4/3}},
% \end{equation}
% where we have denoted the magnetic colatitude of the field line on the NS surface as $\theta=10^{-2}\theta_{-2}\rm\, rad$ (measured from the magnetic pole) and the corresponding curvature radius for a dipolar geometry is $R_{\rm B}\simeq 0.8(R_{\rm ns}/\theta) (R/R_{\rm ns})^{1/2}$.  \note{[$\theta=10^{-2}$ is too narrow to be consistent with rotational sweeping of the beam. Here we should take $\theta=0.1\rm\,rad$ to be the fiducial value.]}

The luminosity is mainly set by the charge starvation radius $R$, and the initial amplitude $\delta B$ as well as the transverse wavelength $\lambda_\perp$ of the Alfv{\'e}n waves. Our poor understanding of the charge density profile of the magnetosphere does not allow us to directly determine $R$. Generally, Alfv{\'e}n waves launched near the magnetic poles where field lines extend to large distances are much more likely to become charge starved and produce coherent radiation.
% We cannot determine $R$ directly by the poorly understood density profile of the magnetosphere.
Here, we can use observed luminosity of FRB 200428, $L_{\rm frb}\sim 3\times 10^{38}\rm\, erg\,s^{-1}$ (assuming a distance of $\sim$$10\rm\, kpc$ and frequency bandwidth of $\Delta \nu \sim 1.4\rm\, GHz$), to constrain $R/R_{\rm ns}\sim 20\, (\delta B_{10}/\lambda_{\perp4})^{6/11}$.
% A self-consistent but non-unique solution is as follows: the Alfv{\'e}n wave amplitude $\delta B\sim 10^{9}\rm\, G$, transverse wavelength $\lambda_{\perp}\sim 10^4{\rm\, cm}$, and charge starvation (or FRB emission) radius $R/R_{\rm ns}\sim 5$. 

The spectrum of the emergent radio waves depends on the size distribution of particle clumps and their Lorentz factors. The emergent power at frequency $\nu$ depends on the Fourier transform of particle number density $\tilde {n}(k)$ at wave number $k\sim 2\pi\nu/c$, and the distribution of particle Lorentz factor ($\gamma$) on this scale.
% The wavelength of curvature radiation is $\sim R_B/\gamma^3$, so if the Lorentz factor of a clump of size $\sim c/\nu$ is far from $(R_B\nu/c)^{1/3}$, there would be very little power produced at this frequency.
% Therefore, the model described here suggests that the spectrum can have large intrinsic variations over narrow band of $\Delta\nu/\nu < 1$.
We note that the transverse size of the coherent patch $\ell_\perp$ is typically much smaller than the causal length $R/\gamma$, which means that Doppler effect only slightly broadens the spectrum by $\Delta \nu/\nu \simeq (\gamma\ell_\perp/R)^2\sim 0.1$, and the spectrum can have large intrinsic variations over a narrow band as radiation arrives from different clumps at different observer time.
% We also note that the transverse size of the coherent patch $\ell_\perp$ is typically much larger than $\gamma\lambda_{\rm frb}$, which means that the half beaming angle of each coherent patch, corresponding to the angular size of the Airy disk $\alpha \simeq 0.6\lambda_{\rm frb}/2\ell_\perp\simeq 10^{-3}R_7^{-1/2}\nu_9^{-1/2}\rm\, cm$, is much smaller than $1/\gamma$. Thus, Doppler effect only slightly broadens the spectrum by $\Delta \nu/\nu \simeq (\gamma \alpha)^2\sim 10^{-2}$.

% The scenario we have considered is that a trigger for an outburst gets started at some depth in the crust of a NS, and then the disturbance propagates to other parts of the star and to the surface where it transfers energy to magnetic fields. The Alfv{\'e}n waves launched near the magnetic poles where field lines extend to large distances are much more likely to become charge starved and produce coherent radiation.

The FRB emission is produced at a radius $R\sim 20 R_{\rm ns}$, as described above, with an uncertainty by a factor of a few. Thus, Alfv{\'e}n waves should be launched within the polar angle $\theta\lesssim 0.1\rm\,rad$ in order to ensure that these waves are able to propagate out to $\sim 10^2 R_{\rm ns}$ and pass through charge starvation point. Furthermore, $\theta$ cannot be much smaller than $0.02\rm\, rad$ because otherwise the beaming cone of field lines at $\sim 20 R_{\rm ns}$ would rotate outside observer line of sight in 30 ms\footnote{It might be tempting to consider the possibility that the two radio pulses separated by 30 ms were in fact due to one continuous event that produced a hollow cone of radio emission, and the two pulses corresponded to the sweep of the cone across the line of sight as the NS rotated. However, two hard X-ray pulses also separated by $\sim 30$ ms cast doubt on this possibility, since it requires that the hard X-ray emission is also beamed into the same hollow cone as the radio emission whereas the softer X-rays were presumably not beamed.} and the second radio pulse seen from FRB 200428 would have been missed. These constraints on the magnetic colatitude motivates the choice of $\theta=0.03\rm\, rad$ as our fiducial value in eq. (\ref{eq:Lfrb}). All things being similar for different X-ray bursts from SGR 1935+2154, we expect to see one FRB for $\sim10^2$ X-ray bursts \citep[this seems consistent with the available data for this object,][]{lin20b}. If the Alfv{\'e}n wave packet has an azimuthal angular span of $\delta \phi \sim 1\rm\, rad$, then the solid angle of FRB emission at the charge starvation radius is $\Omega_{\rm frb} \sim \delta \phi\, \theta^2 (R/R_{\rm ns}) \sim 10^{-2}\rm\, sr$. The beaming fraction of $\Omega_{\rm frb}/4\pi\sim 10^{-3}$ is consistent with that inferred from the volumetric rate of X-ray bursts and FRBs in \S\ref{sec:link}.

In Fig. \ref{fig:constraints}, we show the solutions for different FRBs with a wide range of luminosities, along with a number of physical constraints on the charge starvation radius and the initial amplitude of Alfv{\'e}n disturbance. For simplicity, we fix  $B_{\rm ns}=3\times 10^{14}\rm\,G$, $P=3\rm\,s$, $\theta=10^{-1.5}\rm\,rad$, and $\nu_{\rm frb}=1 \rm\,GHz$. The biggest uncertainty lies on $\lambda_\perp$, the transverse wavelength of the Alfv{\'e}n waves on the NS surface, which depends on how the initial disturbance is launched. For $\lambda_\perp = 10^{4}\rm\, cm$, our model predicts FRB luminosities in the range $10^{36}$ to $10^{48}\rm\, erg\, s^{-1}$ and hence provides a viable explanation for faint bursts like \frb as well as bright events like FRB 190523 \citep{ravi19}. The maximum luminosity is due to the wave electric field, parallel to the magnetic field, at the charge starvation radius exceeding the Schwinger limit \citep{lu19_Lmax}. We also predict that the FRB luminosity function must have a (so-far unobserved) flattening at the lower end, although the exact minimum luminosity $L_{\rm min}$ depends on the unknown $\lambda_\perp$. This is because, for very small initial Alfv{\'e}n amplitude, charge starvation occurs far away from the NS surface where the plasma frequency is below the GHz band, and in this case all charge clumps have longitudinal sizes $\ell_\para> \lambda_{\rm frb}$ and hence the coherent emission at GHz frequencies is strongly suppressed. When the line of sight is outside the beaming cone of angular size $\sim\gamma^{-1}$, the observed luminosity is heavily suppressed by relativistic effects and hence may be below $L_{\rm min}$, but the chance of detection is very small.

What fraction of energy in this event reached near the magnetic poles and contributed to FRB emission? Suppose initially the outburst started far away from the magnetic pole, since most free energy in the tangled magnetic fields is near the equator \citep{thompson02, gourgouliatos16}. Crustal deformations during the flare excite seismic oscillations, preferentially toroidal shear modes which preserve the shape of the star \citep{duncan98, piro05}, and the disturbance propagates along the crust to other parts of the star. Due to the small thickness of the crust $h\sim 0.5$--$1\rm\, km$, the wave undergoes many reflections off the surface before reaching the polar region. The distance traveled by the wave between two consecutive reflections is $\ell \sim \sqrt{hR_{\rm ns}}$, and the minimum number of reflections between the trigger to the magnetic pole is $\pi R_{ns}/2\ell\sim 4$. The FRB duration is given by propagation delay between different paths $t_{\rm frb}\sim \ell/v_{\rm s}\sim 1\rm\, ms$ for wave speed $v_{\rm s}\sim 0.01c$. Each time the waves reach the surface, high-frequency ($\gg 10^4\rm\, Hz$) Fourier components is largely transmitted into magnetospheric Alfv{\'e}n modes \citep{blaes89}. The Alfv{\'e}n waves launched at $\theta \gtrsim 0.1\rm\,rad$ are trapped in closed field lines, cascade to smaller scales, and create an $e^\pm$-photon plasma that radiates most of the energy as X-rays. For low-frequency seismic components $\lesssim 10^4\rm\, Hz$, since the corresponding Alfv{\'e}n wavelength is $\gtrsim 3R_{\rm ns}$, their transmission to the magnetosphere preferentially occurs near the poles where the magnetic field lines are sufficiently extended \citep{thompson95}. The energy per unit surface area transmitted into the magnetospheric Alfv{\'e}n waves in the polar region can be estimated to be $F_{\rm A}/F\sim T(1-T)^{N_{\rm r}} h/R_{\rm ns}$, where the fluence normalization $F = E/4\pi R_{\rm ns}^2$ is from uniformly distributing the total energy over the NS surface, $T$ is the transmission coefficient from crustal shear waves to Alfv{\'e}n waves, and $N_{\rm r}\sim 5$ is the typical number of reflections. The frequency spectrum of seismic oscillations and their propagation properties are still highly uncertain. For $0.03\lesssim T\lesssim 0.5$ \citep{blaes89, bransgrove20}, we roughly estimate $F_{\rm A}/F$ to be of order $10^{-3}$. Following the field lines from the NS surface to the charge starvation radius, the energy per solid angle drops by another factor of $R_{\rm ns}/R\sim 0.1$. Assuming a fraction of order unity of the Alfv{\'e}n luminosity is converted into coherent radio emission, we expect the FRB to X-ray luminosity ratio to be of order $10^{-4}$, which is in rough agreement with observed fluence ratio between these two bands.

\section{Summary}\label{sec:summary}
The first Galactic FRB from a magnetar, with its associated X-ray counterpart, provides an extraordinary opportunity to understand the FRB phenomenon as a whole. We explore the implications of \frb in various aspects. We find that FRB 200428-like events likely contribute a significant fraction of the cosmological FRB rate function at the faint end near specific energy $E_\nu\sim 10^{26}\rm\, erg\, Hz^{-1}$. We compared \sgr with the sources of other active repeaters (e.g., FRB 121102) and discuss how they may be understood in a general framework of the magnetar progenitors from different formation channels. Then, we compare the rates of SGR X-ray bursts and FRBs and find that only a small fraction (of order $10^{-3}$--$10^{-2}$) of X-ray bursts may be accompanied by FRBs.

We consider two broad classes of FRB emission mechanisms. First, the \tq{far-away} models describe that a relativistic outflow drives a shock into the surrounding medium at large distances and generates radio emission by a plasma maser process. We carried out a detailed analysis of these models and found a number of difficulties explaining the radio and X-ray data from FRB 200428. The second class are the \tq{close-in} models where radio emission is generated by a coherent process within the NS magnetosphere. We propose a scenario that magnetic disturbance near the stellar surface propagates to larger radii in the form of Alfv{\'e}n waves which then damp and produce radio emission. FRB 200428 was associated with an X-ray burst, and the hard X-ray lightcurve had two prominent spikes that occurred at nearly the same time as the two FRB pulses. The coincidence of hard X-ray and radio peaks and their relative fluxes can be understood in this scenario. This model provides a unified picture for faint bursts like \frb as well as the bright events seen at cosmological distances.

\section{Data Availability}
The data underlying this article will be shared on reasonable request to the corresponding author.

\section{acknowledgments}
WL thank Shri Kulkarni for his encouragement throughout this project. This work has been funded in part by an NSF grant AST-2009619. WL was supported by the David and Ellen Lee Fellowship at Caltech.

{\small
\bibliographystyle{mnras}
\bibliography{refs}
}

\appendix
\section{``Far-away'' models --- Emission from beyond the light cylinder}

In this Appendix, we study the other class of \tq{far-away} models where a relativistic outflow drives a shock into the circum-stellar medium (CSM) at large distances and FRB is generated by a plasma maser process, as proposed by \citet{lyubarsky14, beloborodov17, beloborodov19, metzger19} and further developed by \citet{plotnikov19} and \citet{margalit20}.

% \subsection{Emission from beyond the light cylinder --- synchrotron maser model}\label{sec:shock}
The properties of the CSM may be highly diverse as it is shaped by the pulsar wind, flares from the NS, and the supernova remnant. These complications can be avoided by considering one snapshot in the FRB lightcurve. The observed flux at a given time can be shown to be produced when the shock front is at some effective radius $r$. We take the average density of the material swept up by the shock front up to radius $r$ to be $\rho_0$, bulk Lorentz factor of the unshocked, upstream medium, to be $\Gamma_0$, and magnetization parameter $\sigma = 1 + B_0^2/4\pi \rho_0 c^2$; where $B_0$, the magnetic field strength of the upstream fluid, and $\rho_0$ are measured in the comoving frame of the upstream medium. The CSM is initially cold. The ejecta drives a shock into the CSM and the shock Lorentz factor in the lab or NS rest frame is $\Gamma_{\rm s}$.  The energy of the shocked CSM at radius $r$ is
\begin{equation}\label{eq:total_energy}
    E \simeq 4\pi r^3 u_0 \Gamma_{\rm rel}^2 = 4\pi r^3 \rho_0c^2 (\Gams/\Gamma_0)^2,
\end{equation}
where we have used $u_0 = \sigma \rho_0 c^2$ as the average energy density of unshocked CSM up to radius $r$  and the relative Lorentz factor between the shocked and pre-shock plasma $\Gamma_{\rm rel} = \Gams/(\Gamma_0 \sigma^{1/2})$ as given by the Rankine-Hugoniot jump conditions \citep[e.g.,][]{kennel84}. The emission frequency of the maser emission $\omega=2\pi \nu$ is roughly given by \citep{plotnikov19}
\begin{equation}\label{eq:frb_frequency}
    \omega \simeq 3\Gams \omp,
\end{equation}
where $\omp=\sqrt{4\pi n_0 q^2/\me}$ is the plasma frequency, $n_0=\rho_0/m$ is the electron number density of the upstream plasma, and $m$ is the mean mass per electron.
% \footnote{\note{[do not agree]}\bf The consideration of an electron-proton CSM is the best case scenario for the most efficient generation of coherent radio waves. The blast wave energy increases by several orders of magnitude if the medium upstream of the shock front is considered to be electron-positron.}.
The emission duration $t_{\rm frb}$ is given by\footnote{From pressure balance between the shocked regions, one obtains the relative Lorentz factor $\Gamma_{\rm rel}\simeq (L/L_0)^{1/4}$, where $L = E/t_{\rm ej}$ is the luminosity of the ejecta, $t_{\rm ej}$ is the launching duration, and $L_0 = 4\pi r^2 u_0 \Gamma_0^2 c$ is the luminosity of the outflowing CSM. This combined with eq. (\ref{eq:shock_radius}) then gives $t_{\rm frb}\simeq t_{\rm ej}/(2\sigma)$ \citep{kumar15}, which means the FRB duration is much shorter than the ejection duration if $\sigma\gg 1$.}
\begin{equation}\label{eq:shock_radius}
    r\simeq 2\Gams^2 t_{\rm frb}c.
\end{equation}

\begin{figure}
\centering
\includegraphics[width=0.47\textwidth]{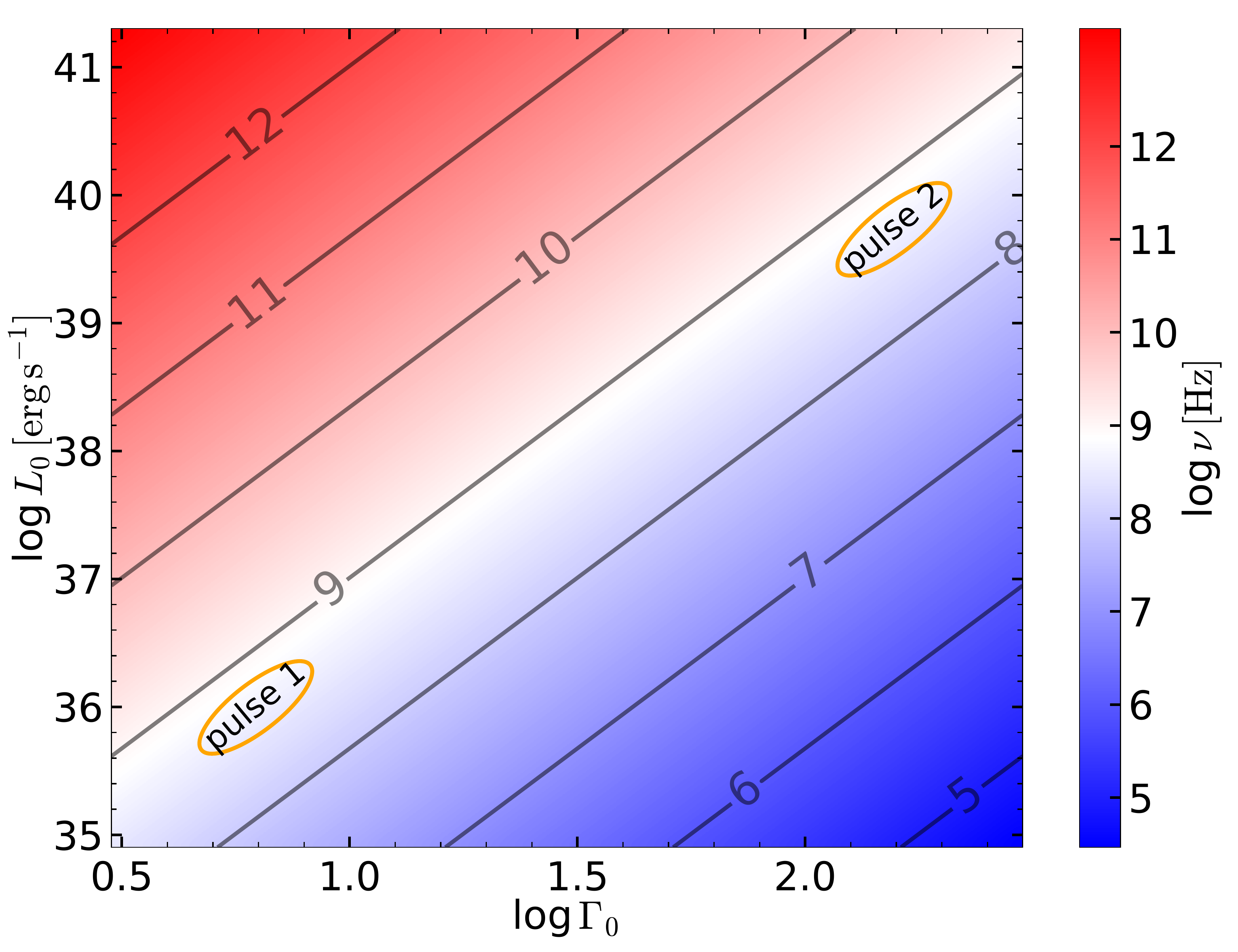}
\caption{This graph (with black contour lines) shows the FRB frequency $\nu$ as a function of the upstream plasma luminosity $L_0$ and Lorentz factor $\Gamma_0$ for the shock-maser model described in this Appendix. We see that, to generate GHz radio emission, the upstream plasma parameters must lie in a very narrow range of the allowed space (along $L_0\propto \Gamma_0^{8/3}$ line). In reality, the physical properties of the CSM (that the relativistic, magnetar flare, ejecta runs into) is expected to be highly diverse as it is shaped by many different processes, and the probability that the maser emission from the shock falls in the observing band is extremely small (the figure shows that the frequency of the emergent maser emission could be anywhere between 10$^6\,$Hz and 10$^{12}\,$Hz for the parameters of FRB 200428 according to the shock model). We also show the locations of the two radio pulses as observed by CHIME by orange ellipses. For this plot, we fix the upstream magnetization $\sigma=1$, ejecta energy $E=10^{40}\rm\, erg$, and FRB duration $t_{\rm frb}=1\rm\, ms$, as motivated by FRB 200428.
}
\label{fig:L0Gam0}
\end{figure}

It is straight-forward to solve the above three equations for the emission radius $r$, shock Lorentz factor $\Gams$, and pre-shock number density $n_0$. And we find
\begin{equation}\label{eq:solution}
\begin{split}
    % r &\simeq (7.3\times10^{10}\mr{\,cm})\, \left(m\over \mp\right)^{-{1\over 3}}\Gamma_0^{2\over 3} E_{40}^{1\over 3} \nu_9^{-{2\over 3}},\\
    % \Gams &\simeq 35\, \left(m\over \mp\right)^{-{1\over 6}}\Gamma_0^{1\over 3} E_{40}^{1\over 6}  \nu_9^{-{1\over 3}} t_{\rm frb,ms}^{-{1\over 2}},\\
    % n_0 &\simeq (1.1\times10^6 \mr{\,cm^{-3}})\, \left(m\over \mp\right)^{1\over 3}\Gamma_0^{-{2\over 3}} E_{40}^{-{1\over 3}}  \nu_9^{8\over 3} t_{\rm frb,ms}.
    r &\simeq (8.9\times10^{11}\mr{\,cm})\, \left(m\over \me\right)^{-{1\over 3}}\Gamma_0^{2\over 3} E_{40}^{1\over 3} \nu_9^{-{2\over 3}},\\
    \Gams &\simeq 1.2\times10^2\, \left(m\over \me\right)^{-{1\over 6}}\Gamma_0^{1\over 3} E_{40}^{1\over 6}  \nu_9^{-{1\over 3}} t_{\rm frb,ms}^{-{1\over 2}},\\
    n_0 &\simeq (9.0\times10^4 \mr{\,cm^{-3}})\, \left(m\over \me\right)^{1\over 3}\Gamma_0^{-{2\over 3}} E_{40}^{-{1\over 3}}  \nu_9^{8\over 3} t_{\rm frb,ms}.
\end{split}
\end{equation}
The optical depth of the upstream plasma for induced Compton (IC) scattering is given by \citep[e.g.,][]{lyubarsky08, kumar20a}
\begin{equation}
    % \tau_{\rm IC} \simeq {3\sigma_{\rm T} E_{\rm frb} \Gamma_0^2 n_0 c \over 32\pi^2 r^2 \me \nu^3} \simeq 4.3\times10^3 {m\over \mp} f_{\rm r,-5}\, \nu_9\, t_{\rm frb,ms}.
    \tau_{\rm IC} \simeq {3\sigma_{\rm T} E_{\rm frb} \Gamma_0^2 n_0 c \over 32\pi^2 r^2 \me \nu^3} \simeq 23 {m\over \me} f_{\rm r,-4}\, \nu_9\, t_{\rm frb,ms}.
\end{equation}
To allow GHz coherent radio waves to escape, we find that the upstream composition must be dominated by electron-positron pairs\footnote{For a baryonic (electron-proton) composition, the radiative efficiency must be extremely low $f_{\rm r}\lesssim 10^{-7}$ in order to have $\tau_{\rm IC}\lesssim 10$. And that requires an ejecta energy of $E\gtrsim 10^{43}\rm\, erg$, which is 3 orders of magnitude higher than seen in the associated hard X-ray burst.} with $m\simeq \me$. In fact, the baryonic shock model is ruled out by the data since it overproduces X-ray luminosity by a factor $\gtrsim10^3$. This is because the baryonic shock must have much larger energy to get around the induced Compton constraints. Hereafter, we take $m=\me$ and then the luminosity of upstream material is
\begin{equation}\label{eq:L_upstream}
    L_0 \simeq (2.2\times10^{34}\mr{\,erg\,s^{-1}})\, \sigma \Gamma_0^{8\over 3} E_{40}^{1\over 3} \nu_9^{4\over 3} t_{\rm frb,ms}.
\end{equation}
In Fig. \ref{fig:L0Gam0}, we show how the FRB frequency is related to the upstream luminosity $L_0$ and Lorentz factor $\Gamma_0$ according to the above relation, while fixing $E_{40}=1$ and $t_{\rm frb,ms}=1$ as motivated by FRB 200428. We see that, to generate GHz radio emission, the upstream plasma conditions must lie along a narrow valley in the otherwise very wide parameter space. 

The next step is to consider that there are two radio pulses separated by about 30 ms as detected by CHIME. The first one spans from 400 MHz (lower end of the observing band) up to 550 MHz, and the second one spans from 550 to 800 MHz (upper end of the observing band). One should be cautious about the details of the spectrum because \frb is detected in the far side lobe where the spectral response may not be well understood. However, since CHIME's response is not expected to change significantly on a timescale of 30 ms, the \textit{difference} between the spectra of the two pulses should be physical. Each pulse has duration of about 1 ms, after correcting for scattering broadening. We also note that the associated X-ray burst also had two distinct peaks separated by 30 ms in the hardest band (27-250 keV) of HXMT \citep{li20_hxmt}, which were temporally coincident with the two radio peaks. This suggests that the two radio pulses are generated by two separated ejectas. The first ejecta interacts with the (perhaps temporarily enhanced) magnetar wind. The second ejecta interacts with the slower tail of the first ejecta or the magnetar wind in between the two flare ejectas responsible for the two radio pulses.

The second ejecta will catch up with the tail of the previous ejecta or the wind following the previous ejecta, which we take to be moving with Lorentz factor $\Gamma_{\rm t}$, at $\delta t\simeq 30\rm\, ms$ in the observer's frame, at the radius
\begin{equation}\label{eq:catch_up_radius}
    r\simeq 2\Gamma_{\rm t}^2 \delta t\, c\simeq 1.8\times10^{9}\mr{\, cm} \, \Gamma_{\rm t}^2.
\end{equation}
We combine eq. (\ref{eq:catch_up_radius}) with the expressions in eq. (\ref{eq:solution}) to obtain
% solve for the emission radius, shock Lorentz factor, upstream number density, and upstream Lorentz factor,
\begin{equation}\label{eq:full_solution}
\begin{split}
    r &\simeq (2.0\times10^{13}\mr{\,cm})\, E_{40}^{1\over 2} \nu_9^{-1},\\
    \Gams &\simeq 5.7\times10^2\, E_{40}^{1\over 4}  \nu_9^{-{1\over 2}} t_{\rm frb,ms}^{-{1\over 2}},\\
    n_{\rm t} &\simeq (4.0\times10^3 \mr{\,cm^{-3}})\, E_{40}^{-{1\over 2}}  \nu_9^{3} t_{\rm frb,ms},\\
    \Gamt &\simeq 105\, E_{40}^{1/4} \nu_9^{-{1\over 2}},
\end{split}
\end{equation}
where $n_{\rm t}$ is number density of the upstream plasma in its comoving frame. We see that the dynamics of the second ejecta is well determined\footnote{In fact, the dynamics of the first ejecta can also be determined if we assume the density profile of the upstream plasma ahead of the first shock to be $n_0\propto r^{-2}$ (or other power-law forms). This is because, at the observer's time $t\simeq 1\rm\, ms$ (during the first radio pulse of FRB 200428), the first ejecta is at its deceleration radius, which is a factor of $30^{1/2}$ less than the first expression in eq. (\ref{eq:full_solution}). Then, one can plug the deceleration radius back into eq. (\ref{eq:solution}) to solve for the unknown $\Gamma_0$ and hence other quantities as well.}, thanks to the resolved X-ray lightcurve by HXMT. The upstream Lorentz factor $\Gamt$ is reasonable if the first ejecta has most of the energy near the front end with high Lorentz factor $\gg 100$ (which is responsible for the first radio pulse) and a small fraction of energy in the tail with relatively low Lorentz factor $\sim 100$ (which is responsible for decelerating the second ejecta and hence generate the second radio pulse).
% In fact, the dynamics of the first ejecta can also be determined if we assume the density profile of the upstream plasma ahead of the first shock to be $n_0\propto r^{-2}$ (or other power-law forms).
% Then, one can show that, at the observer's time $t\simeq 5\rm\, ms$ (during the first radio pulse of FRB 200428), the upstream plasma before the first shock is moving at Lorentz factor $\Gamma_0\simeq 28$ and had comoving number density $n_0\simeq (9.8\times10^{3}\mr{\,cm^{-3}})\,$. 

Can these shocks produce the non-thermal hard X-rays observed by HXMT and other instruments? The answer turns out to be no. The reason is that the characteristic synchrotron frequencies ($\nu_{\rm m}$) for an electron-positron CSM, for the parameters of the two shocks we determined above, are of order 10$^{13}\,$Hz and $10^{15}\,$Hz respectively, much smaller than X-ray frequencies. Simulations suggest that shocks in a magnetized pair plasma might not produce an extended power-law particle spectrum above the average energy per particle \citep{sironi15}, i.e., little emission above $\nu_{\rm m}$. Even ignoring this difficulty, let us assume that the Fermi acceleration operates in the $e^\pm$ magnetized shock and produces power-law particle distribution with index $p\simeq 2$. The emergent synchrotron spectrum then is $F_\nu\propto \nu^{-0.5}$, which is consistent with the observed soft X-ray power-law. The spectrum should extend with the same slope up to $\sim$100 MeV for the shock parameters of eqs. (\ref{eq:solution}) and (\ref{eq:full_solution}). However, Konus-Wind detected no significant emission above 250 keV from this event \citep{ridnaia20}, which suggests that the hard X-rays did not originate in these shocks. 
% However, Konus-Wind detected no significant emission above 250 keV from this event \citep{ridnaia20}, whereas we
% The spectrum should extend with the same slope up to the 
% the synchrotron cooling frequency which can be easily shown to be $\sim10^2$ MeV for the shock parameters of equations (\ref{eq:solution}) \& (\ref{eq:full_solution}). However, Konus-Wind detected no significant emission above 250 keV from this event \citep{ridnaia20}, which suggests that the hard X-rays did not originate in these shocks.

% Twenty nine X-ray outbursts from SGR 1935+2154 were followed by the FAST concurrently with the X-ray observations, but no radio signal was detected down to fluence limit of mJy ms. Many of these X-ray bursts were more luminous than the one associated with FRB 200428. 

In the following, we point out several problems with the shock model.

Since the two radio components have similar frequencies and durations to within factors of order unity and the upstream magnetization is modest $\sigma\lesssim 2$ (otherwise the FRB duration will be much less than a few ms), we infer that the upstream plasma for both shocks must have similar ratio of $L_0/\Gamma_0^{8/3}$ as given by eq. (\ref{eq:L_upstream}). This poses a problem for this model because the physical conditions of the upstream plasma before the two shocks are largely unrelated. The ratio $L_0/\Gamma_0^{8/3}$ could change by many orders of magnitude from one pulse to another in an FRB event -- especially considering that the shock is being driven into the tail end of the previous outburst, or outflow preceding the current flare, which contains a tiny fraction of the total energy of the outburst -- and the resulting synchrotron maser emission will generally be at widely separated frequencies and produce pulses of very different durations in the observer frame. For instance, if $L_0$ were to be different for the two shocks by a factor 2, then the maser frequency in the observer frame would be different by a factor 1.7 (for the same pulse duration). A factor of 2 change in $\Gamma_0$ would lead to a factor 4 change in the maser frequency. The same argument applies to other close burst pairs such as the ones seen in FRB 121102 \citep{hardy17}.

The typical variability time of the emission from a relativistic shock should be of order the signal arrival time in the observer's frame, $\Delta t\sim t$, because the observed flux at a given moment comes from a wide range of emitting radii of $\Delta r\sim r$ and angles with respect to the line of sight $\theta\sim 1/\Gams$. However, the de-dispersed lightcurves of the two pulses in \frb show extremely rapid rise with $\Delta t/t \equiv \xi\sim 0.1$. Some other FRBs also show very rapid variability time as short as tens of micro-seconds \citep{cho20}. An external shock can account for this sharp rise time, provided that the observed flux is produced in a very small emission area $A \sim \xi^2(r/\Gams)^2$, which is much smaller than the size of the causally connected region $r/\Gams$. However, in this case the blastwave energy should be larger by a factor $\xi^{-2}\sim 10^2$ to account for the observed FRB flux. Then, the efficiency decreases to $f_{\rm r}\sim 10^{-7}$, and the energy required in the relativistic shock is $\sim 10^2$ larger than seen in the X-ray band. Furthermore, an even more serious problem is the requirement that the size of the emission patch in the two completely unrelated shocks should be nearly of the same area and similar location wrt. the observer line of sight in order that the observe flux of the two pulses and their rise times are similar.

The observed spectrum of the first (or second) pulse cuts off abruptly above (or below) about 550 MHz. This also poses a problem. The spectrum for the maser-in-shock mechanism is expected to be broad with $\Delta \nu\sim \nu$ due to slightly different Doppler shift for different points on the shock surface within an angle $1/\Gamma_{\rm s}$ from the line of sight. Particularly worrisome is the cutoff of the spectrum of the second pulse below 550 MHz. Even if the maser mechanism is terminated suddenly when the shock is at some radius $r$, we will continue to receive radiation for at least a few times $r/(2c\Gams^2)$, which drifts down in frequency as $1/t$ and the flux declines roughly as $1/t^{-2}$ \citep{kumar00}; $t$ is the observer frame time. Therefore, in the shock scenario, it is not possible to cutoff the observed emission below 550 MHz except by invoking some propagation effects, but then that makes it problematic to explain the first pulse which extends down to 400 MHz merely 30 ms earlier.

Another concern, at least for the second radio pulse, is that the predicted downward frequency drift by the shock model is not observed; the observed frequency should decreases with time as the shock decelerates. From the expression of $\Gams$ in eq. (\ref{eq:full_solution}), the observer's time scales as $t\propto \Gams^{-2}\nu^{-1}$ since the blastwave energy is conserved. For two different frequencies $\nu^{(1)}>\nu^{(2)}$, the shock Lorentz factor must satisfy $\Gams^{(1)}>\Gams^{(2)}$, so we obtain $t^{(1)}/t^{(2)}< \nu^{(2)}/\nu^{(1)}$, meaning that the observed frequency evolution is steeper than $t\propto \nu^{-1}$. However, no significant drift is seen for the second pulse between $550$ and $800\,$MHz, despite the fact that a factor of $\gtrsim$1.5 in arrival time difference should be measurable.

We end this Appendix by concluding that the \tq{far-away} shock-maser model does a good job of explaining the radio emission efficiency $f_{\rm r}\sim 10^{-5}$--$10^{-4}$. The radio waves can escape the upstream plasma without being significantly scattered by the induced Compton process, provided that the upstream composition is electron-positron. However, there are a number of serious problems with the model. (1) The two radio pulses are generated by two shocks driven by different ejectas separated by 30 ms, but the frequency and duration of the radio pulses require that the upstream plasma with which these ejectas collide must have almost identical values of $L_0/\Gamma_0^{8/3}$, even though they are expected to be physically unrelated and their values could have been easily different by a factor $\gtrsim 10$. (2) The rapid variability time $\Delta t\ll t$ can be explained by the model by invoking that the flux at a given time only comes from a small patch of size much smaller than the causally connected region, but that decreases the efficiency by another factor of $\sim 10^2$, i.e., the energy requirement for the relativistic ejecta exceeds X-ray emission by a factor $\sim 10^2$ in this case. Furthermore, it will need to invoke an additional uncomfortable assumption that the size of the emitting patch is nearly the same for the two pulses produced by unrelated shocks. (3) The narrow frequency band, particularly in the second pulse (550-800 MHz), is problematic for the emission from a relativistic shock. This is because even if the shock and the maser emission is suddenly turned off at a certain radius, we would continue to see photons of frequency smaller than 550 MHz arriving to us from an angle wrt. to our line of sight just slightly larger than $\Gams^{-1}$ with flux barely a factor 2 smaller than that at 550 MHz; CHIME should have detected the emission down to 400 MHz. (4) The emission from a decelerating shock drifts downwards in frequency with time, but the expected drift is not observed by CHIME.

% \section{Remaining issues to look into}
% \note{What is the typical beaming angle of the radio emission in the shock scenario? Need to look at the confinement problem by considering the Alfv{\'e}n radius. It can be shown that the ejecta launched accompanying an X-ray burst must be beamed in an angle typically wider than that of the open field lines. }
% {\bf The first problem stems from the fact that the solid angle of the external shock surface is expected to a good fraction of 4$\pi$ steradian. Twenty nine  X-ray outbursts from the progenitor of the Galactic FRB, SGR 1935+2154, were followed by the radio telescope FAST concurrently with the X-ray observations. However, no radio signal was detected down to about mJy-ms. Many of these X-ray bursts were more luminous than the ones associated with FRB 200428. This suggests either that only a few percent of X-ray outbursts lead to relativistic outflows and external shocks or that the radio emission is much more highly beamed than the X-rays. The former possibility is hard to reconcile with the data for FRB 200428 where the two MJy radio pulses coincided with two X-ray flares. And the latter possibility is at odds with the expected broad beam nature of external shocks. }

% {\bf(1) Only a few percent of X-ray flares from SGR 1935+2154 had observed radio signal whereas the broad beam nature of the external shock model suggests that the fraction should be of order unity.}

\label{lastpage}
\end{document}